\documentclass[12pt]{article}
\usepackage{amsmath,amssymb}
\usepackage{graphicx}
\newcommand{\eqdef}{\stackrel{\text{def}}{=}}

\newcommand{\n}{\nonumber\\}
\newcommand{\bm}{\boldsymbol}
\newcommand{\ignore}[1]{}
\numberwithin{equation}{section}
\newcommand{\Romannumeral}[1]{\uppercase\expandafter{\romannumeral#1}}

\newcommand{\II}{\text{\Romannumeral{2}}}

\newtheorem{theo}{\bf Theorem}[section]

\newtheorem{rema}[theo]{\bf Remark}

\newtheorem{cond}[theo]{\bf Condition}
\newtheorem{prop}[theo]{\bf Proposition}
\newcommand{\cH}{\mathcal{H}}
\newcommand{\cP}{\mathcal{P}}
\newcommand{\cT}{\mathcal{T}}
\newcommand{\cE}{\mathcal{E}}

\allowdisplaybreaks[4]

\begin{document}

\baselineskip=20pt

\newcommand{\preprint}{
\vspace*{-20mm}
   \begin{flushright}\normalsize   
  \end{flushright}}
\newcommand{\Title}[1]{{\baselineskip=26pt
  \begin{center} \Large \bf #1 \\ \ \\ \end{center}}}
\newcommand{\Author}{\begin{center}
  \large \bf  Ryu Sasaki\footnote{Electronic mail:\ ryu@yukawa.kyoto-u.ac.jp}\end{center}}
\newcommand{\Address}{\begin{center}
      Department of Physics and Astronomy, Tokyo University of Science,
     Noda 278-8510, Japan
   \end{center}}
\newcommand{\Accepted}[1]{\begin{center}
  {\large \sf #1}\\ \vspace{1mm}{\small \sf Accepted for Publication}
  \end{center}}

\preprint
\thispagestyle{empty}

\Title{Quantum vs classical Markov chains; \\
Exactly solvable examples}

\Author

\Address
\vspace{1cm}

\begin{abstract}
A coinless quantisation procedure of general reversible Markov chains on graphs is presented.
A quantum Hamiltonian $\cH$ is obtained by a similarity transformation of the fundamental 
transition probability matrix $K$ in terms of the square root of the reversible distribution.
The evolution of the classical and quantum Markov chains is described by the solutions of the
eigenvalue problem of the quantum Hamiltonian $\cH$.
About twenty plus exactly solvable Markov chains based on the hypergeometric orthogonal 
polynomials of Askey scheme, derived by Odake-Sasaki,
 would provide a good window for  scrutinising the quantum/classical 
contrast of Markov chains.
Among them five explicit examples, related to the Krawtchouk, Hahn, $q$-Hahn, Charlier and Meixner,
are demonstrated to illustrate the actual calculations.
\end{abstract}

\section{Introduction}
\label{sec:intro}
Quantum versions of random walks on graphs of many varieties, continuous time 
\cite{chase, childs, farhi, kempe, mulken,mulken2, solenov,  xu1, xu2} 
or discrete time with coins of different designs and numbers 
\cite{aharonov, ambainis,  ambainis2, childs2, kempe, konno, nayak,  szegedy,  venegas,watrous} and 
without \cite{meyer, patel}, 
have been a hot topic in many disciplines including computer science and mathematical physics.
In contrast, the quantisation of other well known stochastic processes, {\em e.g.} 
general Markov chains or the birth and death processes, does not seem to have gained much recognition.
The non-symmetric nature of these processes appears to be the main obstacle for undertaking quantisation,
which requires a hermitian or real symmetric Hamiltonian.

In this paper I present a general scheme of coinless quantisation of stationary and {\em reversible}
Markov chains on graphs \cite{aldous}.  The reversibility condition is due to the consistency 
with the Schr\"odinger equations
which are time reversal invariant. A real symmetric Hamiltonian $\cH$ is defined by a similarity transformation of the 
fundamental transition probability matrix $K$ in terms of the square root of the reversible distribution.
The solutions of the quantum and classical time evolution problems are reduced to the eigenvalue problem
of the quantum Hamiltonian $\cH$. 

Detailed comparison of the classical and the newly derived quantum version of the Markov chains 
would reveal the merits/demerits of the quantum calculus. 
More than twenty explicit examples of exactly solvable Markov chains  based on the
hypergeometric orthogonal polynomials of Askey scheme \cite{askey, ismail,kls} derived by Odake and myself 
about a few years ago \cite{os39} are available for this purpose.
The analytic expressions of all the eigenvalues and eigenvectors of the transition probability 
matrix $K$ and the Hamiltonian $\cH$ are ready to use.
I pick up five typical examples to illustrate the actual calculations.
The outline of the present paper is similar to that of a previous paper offering the coinless quantisation
of continuous and discrete time birth and death processes \cite{dtbd,qcbd}.
This paper is prepared in a plain style so that non-experts can easily understand.

The present  paper is organised as follows. In section two the problem setting of classical Markov chains is recapitulated 
in order to introduce necessary notions and notation. 
Given the fundamental transition probability matrix $K$ and the reversible distribution $\pi(x)$,
the corresponding quantum Hamiltonian $\cH$  is introduced in {\bf Proposition \ref{pro5}}.
The evolution of Markov chains is summarised in two theorems in section three;
the classical evolution in {\bf Theorem \ref{theo1}} and the quantum one in {\bf Theorem \ref{theo2}}.
In section four the main claim of the paper is stated in {\bf Theorem \ref{theo3}}. 
In the rest of the section  the details of 5 explicit examples of exactly solvable Markov chains are supplied. 
Section five is for a few comments.
Basic definitions related to the ($q$)-hypergeometric functions are supplied in Appendix for self-consistency.

\section{Problem setting, classical }
\label{sec:set}
In order to formulate the quantum version of  the general stationary Markov chains on graphs,
let us first recapitulate the well-known problem setting of the classical Markov chains  on 
a graph $G=(V,E)$, \cite{aldous}.
The set of vertices $V$ of the graph $G$ is is either finite or semi-infinite;
\begin{equation}
V=\{0,1,\ldots,N\}: \quad \text{finite},\qquad V
=\mathbb{Z}_{\ge0}: \quad \text {semi-infinite}.
\label{GVx}
\end{equation}
For analytic treatments  I  use $x,y,..$ as representing the the vertices of $V$,  $x,y\in{V}$ 
rather than the conventional $n$ and $m$ etc. The set of edges of $G$ is denoted by $E$.
In quantum mechanics with a Hamiltonian $\cH$, time evolution is usually  denoted by
$\langle{\rm final\ state}|e^{-i\cH t}|{\rm initial\ state}\rangle$. 
Likewise I use the convention that the transition probability matrix per unit time interval $K(x,y)$ on $V$ means
the transition from an initial vertex $y$ to a final vertex $x$. 
In probability community, the opposite convention is dominant.
 The matrix $K$ is non-negative and satisfies the conservation of probability
 \begin{equation}
  K(x,y)\ge0,\quad\sum_{x\in{V}}K(x,y)=1.
  \label{basK}
\end{equation}

The Schr\"odinger equation is time-reversal invariant.
With quantum treatments in mind, the following condition is imposed.
\begin{cond}
\label{cond1}
The Markov chain described by $K$ is connected and reversible. 
That is,  $K$ has a stationary (detailed balanced)
distribution $\pi(x)$ satisfying
\begin{equation}
K(x,y)\pi(y)=K(y,x)\pi(x),\qquad \forall x, y \in{V},\quad \pi(x)>0,\quad \sum_{x\in{V}}\pi(x)=1.
\label{revcond}
\end{equation}
\end{cond}
A simple random walk on a finite connected undirected graph $G=(V,E)$ without a self-loop
becomes reversible for the following choices  \cite{venegas, woess},
\begin{equation}
K(x,y)=
\left\{
\begin{array}{cc}
 \frac1{d(y)} &   \text{if}\ (x,y)\in{E}   \\[4pt]
 0 &   \text{otherwise}
\end{array}
\right.,
\qquad \pi(x)=\frac{d(x)}{2m},
\end{equation}
in which $d(x)$ is the degree of the vertex $x$, {\em i.e.} the number of edges connected to $x$,
 and $m$ is the total number of edges $m=\#(E)$.
Since there are $d(y)$ vertices directly connected to $y$, $\sum_{x\in{V}}K(x,y)=1$ 
and $\sum_{x\in{V}}\pi(x)=1$. The reversibility condition \eqref{revcond} is trivially satisfied.

The right and left eigenvectors of $K$ of eigenvalue 1 are easy to find.
\begin{prop}
\label{pro1}
The right and left eigenvectors of $K$ of eigenvalue 1 are $\pi(x)$ and 1,
\begin{align}
\sum_{y\in{V}}K(x,y)\pi(y)&=\pi(x)\sum_{y\in{V}}K(y,x)=\pi(x),
\label{pieig}\\
\sum_{y\in{V}}K(y,x)v_0(y)&=\sum_{y\in{V}}K(y,x)=1=v_0(x), \quad v_0(x)\equiv1.
\label{1eig}
\end{align}
It should be stressed that  $v_0(x)\equiv1$ is not an eigenvector  when the graph G
is infinite, as it is not  square summable $v_0(x)\notin\ell^2(V)$.
\end{prop}
As a sequel to $v_0(x)$, let us consider the general solutions of the left eigenvalue problem
of $K$,
\begin{equation}
\sum_{y\in{V}}K(y,x)v_n(y)=\kappa(n)v_n(x),\quad n\in{V}, \quad \kappa(0)=1.
\label{leftprob}
\end{equation}

Next let us introduce the square root of the stationary distribution
$\pi(x)$ and a diagonal matrix $\Phi$ on V,
\begin{equation}
  \hat{\phi}_0(x)\eqdef\sqrt{\pi(x)},\quad
  \Phi(x,x)\eqdef\hat{\phi}_0(x),\quad
  \Phi(x,y)\eqdef0\ \ (x\neq y).
  \label{Phidef}
\end{equation}
\begin{prop}
\label{pro2}
By dividing both sides of the reversibility condition \eqref{revcond} by
$\sqrt{\pi(x)\pi(y)}$, a non-negative and  real symmetric  matrix $\cT$ is obtained,
\begin{align}
 & \cT(x,y)\eqdef \frac1{\sqrt{\pi(x)}}K(x,y)\sqrt{\pi(y)}=\frac1{\sqrt{\pi(y)}}K(y,x)\sqrt{\pi(x)}=\cT(y,x),
  \label{realsym}\\[4pt]
& \hspace{3cm} \cT=\Phi^{-1}K\Phi,
 \label{KTsim}\\
 &\Rightarrow \sum_{y\in{V}}\cT(x,y)\sqrt{\pi(y)}v_n(y)
 =\sqrt{\pi(x)}\sum_{y\in{V}}K(y,x)v_n(y)=\kappa(n)\sqrt{\pi(x)}v_n(x),\quad n\in{V},
 \label{Tveq}\\
 &\Rightarrow \bigl(\cT\phi_n\bigr)(x)=\kappa(n)\phi_n(x),\quad 
 \phi_n(x)\eqdef\sqrt{\pi(x)}v_n(x)=\hat{\phi}_0(x)v_n(x),\qquad \qquad \ \, n\in{V},
  \label{Tveq2}\\[2pt]
& \qquad  \bigl(K\hat{\phi}_0\phi_n\bigr)(x)=\sum_{y\in{V}}K(x,y)\pi(y)v_n(y)
=\pi(x)\sum_{y\in{V}}K(y,x)v_n(y)=\kappa(n)\pi(x)v_n(x)\\
&\hspace{3cm} =\kappa(n)\hat{\phi}_0(x)\phi_n(x),
\hspace{7.3cm} n\in{V}.
\label{Kveq}
\end{align}
The non-negative and real  symmetric matrix corresponding to $\cT$ was reported in \cite{aldous}.
\end{prop}
The vectors $\{\phi_n\}$ are eigenvectors of $\cT$ for finite graphs $G$. For infinite graphs they have to be
square summable, $\phi_n\in\ell^2(V)$,
\begin{align} 
(\phi_n,\phi_m)\eqdef \sum_{x\in{V}}\phi_n(x)\phi_m(x)
= \sum_{x\in{V}}\pi(x)v_n(x)v_m(x)
=\frac1{d_n^2}\delta_{n\,m},
 \quad n, m\in{V}.
  \label{orth}
\end{align}
Here the normalisation constants $\{d_n\}$ are positive $d_n>0$.
In case of degenerate eigenvalues, the orthogonalisation is assumed.
For the exactly solvable examples introduced in section four, the orthogonality of the eigenvectors is
 guaranteed.

This leads to the following 
\begin{prop}
\label{pro3}
$K$ and $\cT$ share all the eigenvalues $\{\kappa(n)\}$ \eqref{leftprob} 
which are real and Perron-Frobenius theorem tells
\begin{equation}
  -1\le\kappa(n)\le1,\quad n\in{V},
\label{perron}
\end{equation}
and the eigenvectors are related by the multiplication of $\hat{\phi}_0(x)$, 
\begin{equation}
\bigl(\cT\phi_n\bigr)(x)=\kappa(n)\phi_n(x),\quad \bigl(K\hat{\phi}_0\phi_n\bigr)(x)
=\kappa(n)\hat{\phi}_0(x)\phi_n(x),\quad n\in{V}.
\label{TKphin}
\end{equation}
If $K$ is a positive matrix, $-1$ does not belong to the set of eigenvalues, $-1<\kappa(n)$.
In all the exactly solvable examples to be introduced in section four, $K$ is a positive matrix.
\end{prop}
A candidate for a possible quantum Hamiltonian $\cH$ of the classical Markov chain $K$ 
is introduced in the following.
\begin{prop}
\label{pro4}
A real symmetric and positive semi-definite matrix $\cH$ on the Hilbert space $\ell^2(V)$ is defined by
\begin{align} 
  \cH&\eqdef \bm{1}-\cT=\Phi^{-1}\left(\bm{1}-K\right)\Phi.
  \label{Hdef}
\end{align}
It shares all the eigenvectors with $\cT$,
\begin{align} 
& \bigl(\cH\phi_n\bigr)(x)=\cE(n)\phi_n(x),\qquad 0\leq  \cE(n)=1-\kappa(n)\le2,\quad n\in{V}.
\label{eigH}
\end{align}
\end{prop}
Here $\bm{1}$ is the identity matrix on $\ell^2(V)$. In theories of random walk on infinite graphs 
\cite{woess} $K-\bm{1}$ is the Laplacian 
and the negative Laplacian is a good candidate of a quantum Hamiltonian.
In terms of the orthonormal basis $\{\hat{\phi}_n(x)\}$ of $\ell^2(V)$,
\begin{equation}
\hat{\phi}_n(x)\eqdef\phi_n(x)d_n,\quad n\in{V},
\label{orthdef}
\end{equation}
the spectral representations of $\cH$ and $K$ read as follows,
\begin{prop}
\label{pro5}
\begin{align} 
\cH(x,y)&=\sum_{n\in{V}}\bigl(1-\kappa(n)\bigr)\hat{\phi}_n(x)\hat{\phi}_n(y),
\qquad \qquad x,y\in{V},
\label{Hspec}\\
K(x,y)&=\hat{\phi}_0(x)\hat{\phi}_0(y)^{-1}\sum_{n\in{V}}\kappa(n)\hat{\phi}_n(x)\hat{\phi}_n(y),\quad x,y\in{V}.
\label{Kspec}
\end{align}
\end{prop}
\section{The evolution of Markov chains}
\label{sec:evol}
Some fundamental formulas of the evolution of the reversible Markov chains described by 
$K(x,y)$ \eqref{basK}--\eqref{Kspec} are listed here, classical vs quantum.

\subsection{Classical evolution}
\label{sec:cevol}

Let $\mathcal{P}(x;\ell)\ge0$ ($\sum_{x\in{V}}\mathcal{P}(x;\ell)=1$)
be the probability distribution  over $V$ at  $\ell$-th time step.
The next step distribution  is given by
\begin{equation}
\mathcal{P}(x;\ell+1)=\sum_{y\in{V}}K(x,y)\mathcal{P}(y;\ell),
\label{nextstep}
\end{equation}
 and the condition $\sum_{x\in{V}}K(x,y)=1$ \eqref{basK} guarantees the conservation of probability \\
$\sum_{x\in{V}}\cP(x;\ell+1)=1$
during the evolution.

The initial value problem is to determine $\cP(x;\ell)$  for given $\cP(x;0)$,
\begin{equation}
\cP(x;\ell)=\sum_{y\in{V}}(K)^\ell(x,y)\cP(y;0),\quad \sum_{x\in{V}}\cP(x;0)=1,\quad \cP(x;0)\ge0.
\label{dinipro}
\end{equation}

The solutions are summarised by the following well known 
\begin{theo}
\label{theo1}
In terms of the spectral representation of $K$ \eqref{Kspec} 
the solution of the initial value problem of the classical Markov chain   
is given by
\begin{equation}
\mathcal{P}(x;\ell)=\hat{\phi}_0(x)\sum_{n\in{V}}c_n\bigl(\kappa(n)\bigr)^\ell\hat{\phi}_n(x),
\label{ctbdsol1}
\end{equation}
in which $\{c_n\}$ are determined as the expansion coefficients of the initial distribution $\mathcal{P}(x;0)$,
\begin{equation}
\mathcal{P}(x;0)=\hat{\phi}_0(x)\sum_{n\in{V}}c_n\hat{\phi}_n(x) \Rightarrow
c_0=1,\ c_n=\sum_{x\in{V}}\hat{\phi}_n(x)\hat{\phi}_0(x)^{-1}\mathcal{P}(x;0),
\quad n=1,\ldots.
\label{cndef}
\end{equation}The transition probability from $y$ to $x$ after $\ell$  steps is
\begin{equation}
\mathcal{P}(x,y;\ell)=(K)^\ell(x,y)=\hat{\phi}_0(x)\hat{\phi}_0(y)^{-1}
\sum_{n\in\mathcal{X}}(\kappa(n))^\ell\hat{\phi}_n(x)\hat{\phi}_n(y).
\label{ltrpr}
\end{equation}
If $K$ is positive,  the  eigenvalue $-1$ is absent and  the reversible distribution $\pi(x)$ 
\eqref{revcond} becomes the stationary distribution  after large steps,
\begin{equation}
\lim_{\ell\to\infty}\mathcal{P}(x;\ell)=\pi(x),\qquad \lim_{\ell\to\infty}\mathcal{P}(x,y;\ell)=\pi(x).
\end{equation}
\end{theo}

\subsection{Quantum evolution}
\label{sec:qevol}

The quantum  Markov chain is described in the Hilbert space $\ell^2(V)$ 
with the orthonormal basis
\begin{equation*}
|x\rangle,\quad \langle y|x\rangle=\delta_{y\,x},\quad x,y\in{V}.
\end{equation*}
It is expressed in the conventional vector notation 
\begin{equation}
|x\rangle\equiv e_x=\vspace{-4pt}
\begin{array}{ccccccc}
 & \!\! {0} & {\cdots}& {x}&x+1&{\cdots}& \\
{}^t(  & \!\! 0 &  \cdots& 1& 0&\cdots&)\\
& & & & & &
\end{array}
\ \Longrightarrow \sum_{x\in{V}}|x\rangle\langle x|=\bm{1}.
\label{xcomp}
\vspace{-10pt}
\end{equation}
Another orthonormal basis is the normalised eigenvectors of the Hamiltonian $\mathcal{H}$, 
$\{\hat{\phi}_n\}$, $(\hat{\phi}_m,\hat{\phi}_n)=\delta_{m\,n}$, $n,m\in{V}$ \eqref{orthdef}. 
By using the notation $|x\rangle$, it can be expressed by $\|n\rangle\!\rangle$,
\begin{equation}
 \|n\rangle\!\rangle\eqdef\sum_{x\in{V}}\hat{\phi}_n(x)|x\rangle \ 
 \Longleftrightarrow \
\langle x\|n\rangle\!\rangle=\hat{\phi}_n(x),
\qquad \langle\!\langle m\|n\rangle\!\rangle=\delta_{m\,n},\quad n,m\in{V}.
\end{equation}
Corresponding to \eqref{xcomp} one has 
\begin{equation}
\sum_{n\in{V}}\|n\rangle\!\rangle\langle\!\langle n\|=\bm{1},
\label{ncomp}
\end{equation}
which is the completeness relation
\begin{equation}
\sum_{n\in{V}}\langle x\|n\rangle\!\rangle\langle\!\langle n\|y\rangle
=\sum_{n\in{V}}\hat{\phi}_n(x)\hat{\phi}_n(y)=\delta_{x\,y},\quad x,y\in{V}.
\label{ncomp2}
\end{equation}
The spectral representations of the Hamiltonians $\mathcal{H}$  \eqref{Hspec} in this notation  reads
\begin{align}
\mathcal{H}&=\sum_{n\in{V}}\bigl(1-\kappa(n)\bigr)\|n\rangle\!\rangle\langle\!\langle n\|\ \
\Longleftrightarrow \ \mathcal{H}(x,y)=\langle x|\mathcal{H}|y\rangle\ =\sum_{n\in{V}}\bigl(1-\kappa(n)\bigr)\hat{\phi}_n(x)\hat{\phi}_n(y).
\label{Hspec2}
\end{align}
The evolution of the discrete time system is governed by a unitary matrix $U$ acting on $\ell^2(V)$ 
which is the exponentiation of $\cH$,
\begin{equation}
|\psi(\ell+1)\rangle=U|\psi(\ell)\rangle,\quad U\eqdef e^{-i\cH}
=\sum_{n\in{V}}e^{-i(1-\kappa(n))}\|n\rangle\!\rangle\langle\!\langle n\|\ .
\label{Udef}
\end{equation}

The normalised general initial state $|\psi(0)\rangle$ can be represented as
\begin{align}
|\psi(0)\rangle&=\sum_{z\in{V}}\varphi(z)|z\rangle,\quad \ \varphi(z)\in\mathbb{C},
\quad  \ \sum_{z\in{V}}|\varphi(z)|^2=1.
\label{inic}
\end{align}

Corresponding to {\bf Theorem \ref{theo1}},  the  quantum version is the following
\begin{theo}
\label{theo2}
In terms of  the spectral representation of the quantum Hamiltonian $\cH$ \eqref{Hspec},  the solution of the initial value problem  is given by
\begin{align} 
 |\psi(\ell)\rangle=U^\ell |\psi(0)\rangle
&=\sum_{n\in{V}}e^{-i(1-\kappa(n))\ell}\|n\rangle\!\rangle\langle\!\langle n\|\psi(0)\rangle
\label {inisold} \\
&=\sum_{n,z\in{V}}e^{-i(1-\kappa(n))\ell}\|n\rangle\!\rangle\hat{\phi}_n(z)\varphi(z).
\label {inisol2d}
\end{align}
The time evolution is causal and the non-causal effects of branching appear at the measurements.
For the initial state $|\psi(0)\rangle=|y\rangle$ ($\varphi(z)=\delta_{z\,y}$) 
at $\ell=0$, the probability amplitude of  arriving at 
the state $|x\rangle$ at $\ell$-th step is
\begin{align} 
\Psi(x,y;\ell)\eqdef\langle x|U^\ell|y\rangle&=
\sum_{n\in{V}}e^{-i(1-\kappa(n))\ell}\langle x\|n\rangle\!\rangle\langle\!\langle n\|y\rangle
=\sum_{n\in{V}}e^{-i(1-\kappa(n))\ell}\hat{\phi}_n(x)\hat{\phi}_n(y).
\label{psiformd}
\end{align}
The probability of measuring the state $|x\rangle$ at  $\ell$-th step  is
\begin{align} 
|\Psi(x,y;\ell)|^2&=\!\!
\sum_{n\in{V}}\hat{\phi}_n(x)^2\hat{\phi}_n(y)^2
+2\!\!\sum_{n>m\in{V}}\!\!\cos[\bigl(\kappa(n)\!-\!\kappa(m)\bigr)\ell]
\hat{\phi}_n(x)\hat{\phi}_m(x)\hat{\phi}_n(y)\hat{\phi}_m(y).
\label{psi2}
\end{align}
Various stochastic quantities can be calculated based on these  expressions.
Since pure state quantum mechanics is time-reversal invariant, 
the above expressions do not converge  as $\ell\to\infty$.
In other words, there is no stationary (or terminal, limiting) distribution.
The long time average is 
\begin{align}
\lim_{T\to\infty}\frac1T\sum_{\ell=0}^{T}|\Psi(x,y;\ell)|^2&=\sum_{n\in{V}}\hat{\phi}_n(x)^2\hat{\phi}_n(y)^2,
\label{taverage}
\end{align}
which is symmetric in $x$ and $y$.
\end{theo}
\begin{rema}
\label{contdis}
It should be stressed that in the coinless quantum formulations, the continuous time and the discrete time results 
look almost identical when expressed in terms of the spectral representations.
\end{rema}
\section{Explicitly solvable examples; Quantum cases}
\label{sec:exa}

In \cite{os39} Odake and myself reported about two dozens of reversible classical Markov chains of
finite and infinite dimensions.
They are explicitly solved in the sense all the eigenvalues and the corresponding eigenvectors are 
obtained analytically.
These Markov chains are generated by five different types of convolutions of the orthogonality measures 
\cite{coo-hoa-rah77}
of hypergeometric polynomials of the Askey scheme \cite{askey, ismail,kls,os12}, {\em e.g.}
the Krawtchouk, Hahn and $q$-Hahn polynomials etc.
Some of these Markov chains are generalisations of the earlier works \cite{diaconis20,albert, hoa-rah83}.
Those including $q$-polynomials are totally new.
The $K$ of these solvable Markov chains contain many adjustable free parameters
so that they could enjoy wide ranges of applications in various disciplines. 
The eigenvalues of $K$ are terminated ($q$)-hypergeometric functions, see \eqref{1qHphi}, \eqref{gendiaco}, 
\eqref{3Keig}, \eqref{1Ceig}, \eqref{2Meig} and 
most of them are expressed as monomials of the  powers and ($q$)-shifted factorials of the contained parameters,
see \eqref{1qHphi},  \eqref{3Keig}, \eqref{1Ceig}, \eqref{2Meig}.
The eigenvectors are the ($q$-)orthogonal polynomials themselves.

The main claim  of this paper is the following.
\begin{theo}
\label{theo3}
All of the classical reversible Markov chains reported by Odake and myself {\rm \cite{os39}}
provide exactly solvable quantum Markov chains without a coin.
Since the entire set of the eigenvalues and eigenvectors is available analytically,  
the formulas of {\bf Theorem \ref{theo2}} apply directly.
Various quantities, e.g.  hitting time, mixing time, sampling time, filling time, 
dispersion time etc {\rm \cite{aharonov}}
can be calculated explicitly.
\end{theo}
Below I replicate some typical examples of the solvable Markov chains from \cite{os39} to show
how the formulas in {\bf Theorem \ref{theo2}} look like.
The first three correspond to finite Markov chains and $N$ is the maximal coordinate of the vertices.
The next two are infinite ones.
Throughout this section, the $N$ dependence in the notation of the finite polynomials and the
normalisation constant $d_n$ is suppressed
for simplicity of presentation.

\subsection{Type (i) convolution, $q$-Hahn }
\label{sec:1qHex}
The type (i) convolution for constructing $K$ was employed in (3.1) of \cite{hoa-rah83} 
and (1.1) of \cite{albert} in connection with the `cumulative Bernoulli trials' of Krawtchouk processes.
But the convolution works for other finite polynomials. Here I demonstrate it for the $q$-Hahn polynomials, 
see \S4.1.3 of \cite{os39} for more details.
The fundamental transition probability matrix $K$ for the $q$-Hahn is
\begin{equation}
  K_{qH}(x,y)=\sum_{z=0}^{\min(x,y)}\pi_{qH}(x-z,N-z,b,c)\pi_{qH}(z,y,a,b),\quad x,y\in{V},
  \label{qh13K}
\end{equation}
in which $\pi_{qH}$ is the normalised orthogonality measure of the $q$-Hahn polynomial,
\begin{align}
  &0<\pi_{qH}(x,N,a,b)=\genfrac{[}{]}{0pt}{}{\,N\,}{x}
  \frac{(a\,;q)_x\,(b\,;q)_{N-x}a^{N-x}}{(ab\,;q)_N},\quad
  \genfrac{[}{]}{0pt}{}{\,N\,}{x}\eqdef\frac{(q\,;q)_N}
  {(q\,;q)_x\,(q\,;q)_{N-x}},
  \label{qHpi}
\end{align}
$\sum_{x\in{V}}\pi_{qH}(x,N,a,b)=1$, with the parameter ranges $0<a<1$, $b<1$. 
It is easy to see  the conservation of the probability 
$\sum_{x\in{V}}K_{qH}(x,y)=1$. 
The reversible distribution is $\pi_{qH}(x,N,ab,c)$,
\begin{equation}
  K_{qH}(x,y)\pi_{qH}(y,N,ab,c)=  K_{qH}(y,x)\pi_{qH}(x,N,ab,c).
\end{equation}
It can be verified by using \eqref{qHpi} without performing the summation explicitly.
The eigenvalues and the corresponding eigenvectors are
\begin{align}
& \hspace{3cm}\kappa(n) ={}_3\phi_2\Bigl(
  \genfrac{}{}{0pt}{}{q^{-n},\,abcq^{n-1},\,b}{ab,\,bc}\Bigm|q\,;q\Bigr)
  =\frac{b^n(a\,;q)_n(c\,;q)_n}{(ab\,;q)_n(bc\,;q)_n},%
  \label{1qHphi}\\[2pt]
& \hspace{29mm}
 \check{P}_n(x,ab,c),\\
 & \sum_{y\in{V}}K_{qH}(x,y)\pi_{qH}(y,N,ab,c)\check{P}_n(y,ab,c)
  =\kappa(n)\pi_{qH}(x,N,ab,c)\check{P}_n(x,ab,c).
  \label{KqHsol}
\end{align}  
 The explicit form of the polynomial $\check{P}_n(x,a,b)$ and its  normalisation constant are
\begin{align}
 &  \check{P}_n(x,a,b)=P_n\bigl(\eta(x),a,b\bigr)
  ={}_3\phi_2\Bigl(\genfrac{}{}{0pt}{}{q^{-n},\,abq^{n-1},\,q^{-x}}
  {a,\,q^{-N}}\Bigm|q\,;q\Bigr),\quad \eta(x)=q^{-x}-1,
  \label{qHp}\\[2pt]
 & \sum_{x\in{V}}\pi_{qH}(x,N,a,b)\check{P}_n(x,a,b)\check{P}_m(x,a,b)=\frac1{d_n(a,b)^2}\delta_{n\,m},
 \quad n, m\in{V},
 \label{qHort2}\\
 & \hspace{2.5cm}d_n(a,b)^2=\genfrac{[}{]}{0pt}{}{\,N\,}{n}
  \frac{(a,abq^{-1}\,;q)_n}{(abq^N,b\,;q)_n\,a^n}\frac{1-abq^{2n-1}}{1-abq^{-1}}.
  \label{qHd}
\end{align}
The normalised eigenvectors $\{\hat{\phi}_n(x)\}$ of the Hamiltonian $\cH_{qH}$ are
\begin{align} 
\cH_{qH}(x,y)&=\frac1{\sqrt{\pi_{qH}(x,N,ab,c)}}\left(\bm{1}-K_{qH}(x,y)\right)\sqrt{\pi_{qH}(y,N,ab,c)},\\[2pt]
   \hat{\phi}_n(x)& =\sqrt{\pi_{qH}(x,N,ab,c)}\,d_n(ab,c)\,\check{P}_n(x,ab,c),
\label{qHphin}\\
\bigl(\cH_{qH}\hat{\phi}_n\bigr)(x)&=\sum_{y\in{V}}\cH_{qH}(x,y)\hat{\phi}_n(y)=\bigl(1-\kappa(n)\bigr)\hat{\phi}_n(x).
\end{align}
The probability amplitude $\Psi(x,y;\ell)$ \eqref{psiformd} becomes simple when $y=0$, 
since $\check{P}_n(0)=1$.
\begin{align} 
\Psi(x,0;\ell)&=\sqrt{\pi_{qH}(x,N,ab,c)\pi_{qH}(0,N,ab,c)}
\sum_{n\in{V}}e^{-i(1-\kappa(n))\ell}d_n(ab,c)^2\check{P}_n(x,ab,c),
\label{qHpsiformx0}\\
&\hspace{1cm} \pi_{qH}(0,N,ab,c)=\frac{(c\,;q)_N(ab)^N}{(abc\,;q)_N},\quad  
\pi_{qH}(N,N,ab,c)=\frac{(ab\,;q)_N}{(abc\,;q)_N}.
\end{align}
It is further simplified when $x=N$, as
\begin{equation}
\check{P}_n(N,ab,c)=(-ab)^nq^{n(n-1)/2}\frac{(c\,;q)_n}{(ab\,;q)_n}.
\end{equation}

\subsection{Type (ii) convolution, Hahn }
\label{sec:2Hex}

The type (ii) convolution for the Hahn polynomial is a three parameter generalisation of 
the one parameter model of Burnside process by Diaconis and Zhong \cite{diaconis20}.
The convolution also works for the Krawtchouk polynomial but not for the $q$-Hahn. See \S4.2.2 of \cite{os39} 
for more details.
The fundamental transition probability matrix $K$ for  the Hahn is
\begin{equation}
  K_H(x,y)=\!\!\!\sum_{z=\max(0,x+y-N)}^{\min(x,y)}\!\!\!
  \pi_H(x-z,N-y,b,c)\pi_H(z,y,a,b),
  \label{KH422}
\end{equation}
in which $\pi_{H}$ is the normalised orthogonality measure of the Hahn polynomial,
\begin{align}
  &0<\pi_H(x,N,a,b)=\binom{N}{x}\frac{(a)_x\,(b)_{N-x}}{(a+b)_N},\quad 0<a,b,\quad
   \sum_{x\in{V}}\pi_{H}(x,N,a,b)=1.
    \label{piH}
\end{align}
 It is easy to see  the conservation of the probability 
$\sum_{x\in{V}}K_H(x,y)=1$. 
The reversible distribution is $\pi_{H}(x,N,a+b,b+c)$,
\begin{equation}
  K_{H}(x,y)\pi_{H}(y,N,a+b,b+c)=  K_{H}(y,x)\pi_{H}(x,N,a+b,b+c),
\end{equation}
which  can be verified by using \eqref{piH} without performing the summation explicitly.
The eigenvalues and the corresponding eigenvectors are
\begin{align}
  &\kappa(n)\!=\!{}_3F_2\Bigl(
  \genfrac{}{}{0pt}{}{-n,\,n+a+2b+c-1,\,b}{a+b,\,b+c}\Bigm|1\Bigr)\!\!=\!\!\sum_{k=0}^n\binom{n}{k}(-1)^k\frac{(b)_k(n+a+2b+c-1)_k}{(a+b)_k(b+c)_k},
    \label{gendiaco}\\
& \qquad \qquad \check{P}_n(x,a+b,b+c),\\
  & \sum_{y\in{V}}K_{H}(x,y)\pi_{H}(y,N,a+b,b+c)\check{P}_n(y,a+b,b+c)\n
  &\hspace{2cm}=\kappa(n)\pi_{H}(x,N,a+b,b+c)\check{P}_n(x,a+b,b+c).
  \label{KHsol}
\end{align}
 The explicit form of the polynomial $\check{P}_n(x,a,b)$ and its  normalisation constant are
\begin{align}
  &\check{P}_n(x,a,b)=P_n(x,a,b)
  ={}_3F_2\Bigl(\genfrac{}{}{0pt}{}{-n,\,n+a+b-1,\,-x}{a,\,-N}\Bigm|1\Bigr),
  \label{Hp}\\
  & \sum_{x\in{V}}\pi_{H}(x,N,a,b)\check{P}_n(x,a,b)\check{P}_m(x,a,b)=\frac1{d_n(a,b)^2}\delta_{n\,m},
 \quad n, m\in{V},
 \label{Hort}\\
  &d_n(a,b)^2=\binom{N}{n}\frac{(a)_n\,(2n+a+b-1)(a+b)_N}{(b)_n\,(n+a+b-1)_{N+1}}.
\end{align}
The normalised eigenvectors $\{\hat{\phi}_n(x)\}$ of the Hamiltonian $\cH_{H}$ are
\begin{align} 
\cH_{H}(x,y)&=\frac1{\sqrt{\pi_{H}(x,N,a+b,b+c)}}\left(\bm{1}-K_{H}(x,y)\right)\sqrt{\pi_{H}(y,N,a+b,b+c)},\\
   \hat{\phi}_n(x)& =\sqrt{\pi_{H}(x,N,a+b,b+c)}\,d_n(a+b,b+c)\,\check{P}_n(x,a+b,b+c),
\label{Hphin}\\
\bigl(\cH_{H}\hat{\phi}_n\bigr)(x)&=\sum_{y\in{V}}\cH_{H}(x,y)\hat{\phi}_n(y)=\bigl(1-\kappa(n)\bigr)\hat{\phi}_n(x).
\end{align}
The probability amplitude $\Psi(x,y;\ell)$ \eqref{psiformd} becomes simple when $y=0$, 
since $\check{P}_n(0)=1$,
\begin{align} 
&\Psi(x,0;\ell)=\sqrt{\pi_{H}(x,N,a+b,b+c)\pi_{H}(0,N,a+b,b+c)}\n
&\qquad \qquad \qquad \times\sum_{n\in{V}}e^{-i(1-\kappa(n))\ell}d_n(a+b,b+c)^2\check{P}_n(x,a+b,b+c),
\label{Hpsiformx0}\\
&\pi_{H}(0,N,a+b,b+c)=\frac{(b+c)_N}{(a+2b+c)_N},\quad  
\pi_{H}(N,N,a+b,b+c)=\frac{(a+b)_N}{(a+2b+c)_N}.
\end{align}
It is further simplified when $x=N$, as
\begin{equation}
\check{P}_n(N,a+b,b+c)=(-1)^n\frac{(b+c)_n}{(a+b)_n}.
\end{equation}

\subsection{Type (iii) convolution, Krawtchouk}
\label{sec:3Kex}
The Markov chain generated by  type (iii) convolution was, to the best of my knowledge, 
 first introduced in \cite{os39}.
Here is an example using the Krawtchouk polynomial. 
The convolution is applicable to the Hahn and 
$q$-Hahn polynomials as well. See \S4.3 of \cite{os39} for more details.
The fundamental transition probability matrix $K$ for  the Krawtchouk is
\begin{align}
  K_K(x,y)&=\sum_{z=\max(x,y)}^N\pi_K(x,z,b)\pi_K(z-y,N-y,a),\quad  0<a,b<1,
  \label{KK3}
\end{align}
in which $\pi_{K}$ is the normalised orthogonality measure of the Krawtchouk polynomial,
\begin{align}
  &\pi_K(x,N,a)=\binom{N}{x}a^x(1-a)^{N-x},\quad 0<a<1,\quad
    \sum_{x\in{V}}\pi_{K}(x,N,a)=1.
  \label{piK}
 \end{align}
The reversible distribution is $\pi_{K}(x,N,p)$,
\begin{align}
 & K_{K}(x,y)\pi_{K}(y,N,p)=  K_{K}(y,x)\pi_{K}(x,N,p),\qquad  p\eqdef\frac{ab}{1-b+ab},
 \label{Kpdef}
\end{align}
which  can be verified by using \eqref{piK} without performing the summation explicitly.
The eigenvalues and the corresponding eigenvectors are
\begin{align}  
  &\hspace{3cm}\kappa(n) ={}_1F_0\Bigl(\genfrac{}{}{0pt}{}{-n}{-}\Bigm|abp^{-1}\Bigr)
  =(1-a)^nb^n,\qquad \check{P}_n(x,p),
\label{3Keig}  \\
 & \sum_{y\in{V}}K_{K}(x,y)\pi_{K}(y,N,p)\check{P}_n(y,p)
=\kappa(n)\pi_{K}(x,N,p)\check{P}_n(x,p).
  \label{KHsol2}
\end{align}
The explicit form of the polynomial $\check{P}_n(x,a)$ and its  normalisation constant are
\begin{align} 
  &\check{P}_n(x,a)=P_n(x,a)
  ={}_2F_1\Bigl(\genfrac{}{}{0pt}{}{-n,\,-x}{-N}\Bigm|a^{-1}\Bigr),\quad
  P_n(x,a)=P_x(n,a),
  \label{Kp}\\
  & \sum_{x\in{V}}\pi_{K}(x,N,a)\check{P}_n(x,a)\check{P}_m(x,a)=\frac1{d_n(a)^2}\,\delta_{n\,m},
 \quad n, m\in{V},
 \label{Kort}\\
  &\hspace{2.5cm}d_n(a)^2=\binom{N}{n}\Bigl(\frac{a}{1-a}\Bigr)^n.
\end{align}
The normalised eigenvectors $\{\hat{\phi}_n(x)\}$ of the Hamiltonian $\cH_{K}$ are
\begin{align} 
\cH_{K}(x,y)&=\frac1{\sqrt{\pi_{K}(x,N,p)}}\left(\bm{1}-K_{K}(x,y)\right)\sqrt{\pi_{K}(y,N,p)},\\
   \hat{\phi}_n(x)& =\sqrt{\pi_{K}(x,N,p)}\,d_n(p)\check{P}_n(x,p),
\label{Kphin}\\
\bigl(\cH_{K}\hat{\phi}_n\bigr)(x)&=\sum_{y\in{V}}\cH_{K}(x,y)\hat{\phi}_n(y)=\bigl(1-\kappa(n)\bigr)\hat{\phi}_n(x).
\end{align}
The probability amplitude $\Psi(x,y;\ell)$ \eqref{psiformd} becomes simple when $y=0$, 
since $\check{P}_n(0)=1$,
\begin{align} 
\Psi(x,0;\ell)&=\sqrt{\pi_{K}(x,N,p)\pi_{K}(0,N,p)}\,
\sum_{n\in{V}}e^{-i(1-\kappa(n))\ell}d_n(p)^2\check{P}_n(x,p),
\label{Kpsiformx0}\\
&\pi_{K}(0,N,p)=(1-p)^N,\quad  
\pi_{K}(N,N,p)=p^N.
\end{align}
It is further simplified when $x=N$, as
\begin{align}
&\check{P}_n(N,p)=(-1)^n(p^{-1}-1)^n,\\
&\Longrightarrow
\Psi(N,0;\ell)=\frac{\bigl(ab(1-b)\bigr)^{N/2}}{(1-b+ab)^N}\sum_{n\in{V}}(-1)^ne^{-i(1-(1-a)^nb^n)\ell}\binom{N}{n}.
\end{align}

\subsection{Type (i) convolution, Charlier }
\label{sec:1Cex}
The Charlier polynomial is obtained from the Krawtchouk polynomial by a certain limit 
including $N\to\infty$ \cite{kls}. 
This Markov chain is obtained from that of type (i) Krawtchouk, \S4.1.1 \cite{hoa-rah83}.
The fundamental transition probability matrix $K$ for  the Charlier is, 
\begin{equation}
  K_{C}(x,y)=\sum_{z=0}^{\min(x,y)}\pi_{C}(x-z,b)\pi_{K}(z,y,a),\quad x,y\in{V},
  \quad 0<a, b,
  \label{C11K}
\end{equation}
in which $\pi_{C}$ is the normalised orthogonality measure of the Charlier polynomial,
\begin{align}
  &\pi_C(x,a)=\frac{a^xe^{-a}}{x!},\quad 0<a,\quad \sum_{x\in{V}}\pi_C(x,a)=1.
\label{Cpi}
\end{align}  
The reversible distribution is $\pi_{C}(x,p)$,
\begin{equation}
  K_{C}(x,y)\pi_{C}(y,p)=  K_{C}(y,x)\pi_{C}(x,p),\quad p\eqdef\frac{b}{1-a}.
\end{equation}
It can be verified by using \eqref{Cpi} without performing the summation explicitly.
The eigenvalues and the corresponding eigenvectors are
\begin{align}
& \hspace{3cm}\kappa(n) ={}_1F_0\Bigl(\genfrac{}{}{0pt}{}{-n}{-}\Bigm|bp^{-1}\Bigr)
=a^n,\qquad \check{P}_n(y,p),
   \label{1Ceig}\\
 & \sum_{y\in{V}}K_{C}(x,y)\pi_{C}(y,p)\check{P}_n(y,p)
  =\kappa(n)\pi_{C}(x,p)\check{P}_n(x,p).
  \label{KCsol}
 \end{align}
The explicit form of the polynomial $\check{P}_n(x,a)$ and its  normalisation constant are 
\begin{align}
  &\check{P}_n(x,a)=P_n(x,a)
  ={}_2F_0\Bigl(\genfrac{}{}{0pt}{}{-n,\,-x}{-}\Bigm|-a^{-1}\Bigr),\quad
  P_n(x,a)=P_x(n,a),
  \label{Cp}\\
 & \sum_{x\in{V}}\pi_{C}(x,a)\check{P}_n(x,a)\check{P}_m(x,a)=\frac1{d_n(a)^2}\delta_{n\,m},
 \quad  d_n(a)^2=\frac{a^n}{n!},\quad n, m\in{V}.
 \label{qHort}
\end{align}
The normalised eigenvectors $\{\hat{\phi}_n(x)\}$ of the Hamiltonian $\cH_{C}$ are
\begin{align} 
\cH_{C}(x,y)&=\frac1{\sqrt{\pi_{C}(x,p)}}\bigl(\bm{1}-K_{C}(x,y)\bigr)\sqrt{\pi_{C}(y,p)},\\
   \hat{\phi}_n(x)& =\sqrt{\pi_{C}(x,p)}\,d_n(p)\check{P}_n(x,p),
\label{Cphin}\\
\bigl(\cH_{C}\hat{\phi}_n\bigr)(x)&=\sum_{y\in{V}}\cH_{C}(x,y)\hat{\phi}_n(y)=\bigl(1-\kappa(n)\bigr)\hat{\phi}_n(x).
\end{align}
The probability amplitude $\Psi(x,y;\ell)$ \eqref{psiformd} becomes simple when $y=0$, 
since $\check{P}_n(0)=1$,
\begin{align} 
\Psi(x,0;\ell)&=\sqrt{\pi_{C}(x,p)\pi_{C}(0,p)}\,
\sum_{n\in{V}}e^{-i(1-\kappa(n))\ell}d_n(p)^2\check{P}_n(x,p),
\label{Cpsiformx0}\\
&\pi_{C}(0,p)=e^{-p}.
\end{align}

\subsection{Type (ii) convolution, Meixner }
\label{sec:2Mex}
This Markov chain is obtained from that of Hahn \eqref{KH422} in \S\ref{sec:2Hex} by 
 fixing $a$, $b$ with $c\to N(1-c)c^{-1}$ ($\Rightarrow 0<c<1$) and taking
the limit $N\to\infty$,
\begin{equation}
 K_{M}(x,y)
  =\sum_{z=0}^{\min(x,y)}\pi_{M}(x-z,b,c)\pi_H(z,y,a,b),
  \label{KM422}
\end{equation}
in which $\pi_M$ is the normalised orthogonality measure of the Meixner polynomial,
\begin{equation}
 \pi_M(x,a,b)=\frac{(a)_x\,b^x(1-b)^a}{x!},\quad \sum_{x\in{V}}\pi_M(x,a,b)=1,\quad 
 0<a,\quad 0<b<1.
 \label{Mpi}
\end{equation}
The reversible distribution is $\pi_{M}(x,a+b,c)$,
\begin{equation}
  K_{M}(x,y)\pi_{M}(y,a+b,c)=  K_{M}(y,x)\pi_{M}(x,a+b,c).
\end{equation}
It can be verified by using \eqref{Mpi} without performing the summation explicitly.
The eigenvalues and the corresponding eigenvectors are
\begin{align}
& \hspace{3cm}\kappa(n) ={}_2F_1\Bigl(\genfrac{}{}{0pt}{}{-n,\,b}{a+b}\Bigm|1\Bigr)
=\frac{(a)_n}{(a+b)_n},\qquad \check{P}_n(y,a+b,c),
   \label{2Meig}\\[2pt]
   & \sum_{y\in{V}}K_{M}(x,y)\pi_{M}(y,a+b,c)\check{P}_n(y,a+b,c)
  =\kappa(n)\pi_{M}(x,a+b,c)\check{P}_n(x,a+b,c).
  \label{KMsol}
 \end{align}
 The explicit form of the polynomial $\check{P}_n(x,a,b)$ and its  normalisation constant are 
\begin{align}
  &\check{P}_n(x,a,b)=P_n(x,a,b)
  ={}_2F_1\Bigl(\genfrac{}{}{0pt}{}{-n,\,-x}{a}\Bigm|1-b^{-1}\Bigr),\quad
  P_n(x,a,b)=P_x(n,a,b),
  \label{Mp}\\[2pt]
 & \sum_{x\in{V}}\pi_{M}(x,a,b)\check{P}_n(x,a,b)\check{P}_m(x,a,b)=\frac1{d_n(a,b)^2}\delta_{n\,m},
 \ \ d_n(a,b)^2=\frac{(a)_n\,b^n}{n!},
\ n, m\in{V}.
 \label{Mort}
\end{align}
The normalised eigenvectors $\{\hat{\phi}_n(x)\}$ of the Hamiltonian $\cH_{M}$ are
\begin{align} 
\cH_{M}(x,y)&=\frac1{\sqrt{\pi_{M}(x,a+b,c)}}\left(\bm{1}-K_{M}(x,y)\right)\sqrt{\pi_{M}(y,a+b,c)},\\
   \hat{\phi}_n(x)& =\sqrt{\pi_{M}(x,a+b,c)}\,d_n(a+b,c)\check{P}_n(x,a+b,c),
\label{Mphin}\\
\bigl(\cH_{M}\hat{\phi}_n\bigr)(x)&=\sum_{y\in{V}}\cH_{M}(x,y)\hat{\phi}_n(y)=\bigl(1-\kappa(n)\bigr)\hat{\phi}_n(x).
\end{align}
The probability amplitude $\Psi(x,y;\ell)$ \eqref{psiformd} becomes simple when $y=0$, 
since $\check{P}_n(0)=1$,
\begin{align} 
\Psi(x,0;\ell)&=\sqrt{\pi_{M}(x,a+b,c)\pi_{M}(0,a+b,c)}\n
&\hspace{2cm} \times\sum_{n\in{V}}e^{-i(1-\kappa(n))\ell}d_n(a+b,c)^2\check{P}_n(x,a+b,c),
\label{Mpsiformx0}\\
&\pi_{M}(0,a+b,c)=(1-c)^{a+b}.
\end{align}

\section{Comments}
\label{sec:comm}
A few comments are in order.
\begin{itemize}
\item  I follow the standard convention of quantum mechanics, $\langle {\rm final\ state}|e^{-i\mathcal{H}t}|{\rm initial \ state}\rangle$.
Correspondingly,  the fundamental transition probability matrix $K(x,y)$  means the transition from the initial vertex $y$ 
to the  final vertex $x$.
\item The main point  of the exact solvability of the $K$ constructed by the five types of convolutions shown in \cite{os39}
is that the orthogonal polynomial $\check{P}_n(x)$ is the left eigenvector of $K$,
\begin{equation*}
  \sum_{y\in{V}}K(y,x)\check{P}_n(y)
  =\kappa(n)\check{P}_n(x)\quad (n\in{V}).
  \tag{\cite{os39}.2.12}
\end{equation*}
The proof is given in Appendix of \cite{os39}.
In most cases, the eigenvalue $\kappa(n)$ is known, without calculation,
from the factorised form of the transition probability matrix $K$ due to the convolution.
\item Time averaged distributions of quantum walks were reported in many papers.
It should be stressed that these results are initial state specific and not universal, cf \eqref{taverage}.
\item It should be noted that, because of the context, the present definition of
$d_n^2$ corresponds to $d_n^2/d_0^2$ in the accompanying  paper on the quantum birth and death process 
\cite{qcbd}
 and also previous series of papers
\cite{os12,bdsol,os34}. 
\end{itemize}


\section*{\hspace{7cm}Appendix \\
Basic  definitions related to the  ($q$)-hypergeometric functions}
\label{appendA}
\setcounter{equation}{0}
\renewcommand{\theequation}{A.\arabic{equation}}


\noindent
$\circ$ shifted factorial $(a)_n$ :
\begin{equation}
  (a)_n\eqdef\prod_{k=1}^n(a+k-1)=a(a+1)\cdots(a+n-1)
  =\frac{\Gamma(a+n)}{\Gamma(a)}.
  \label{defPoch}
\end{equation}
$\circ$ $q$-shifted factorial $(a\,;q)_n$ :
\begin{equation}
  (a\,;q)_n\eqdef\prod_{k=1}^n(1-aq^{k-1})=(1-a)(1-aq)\cdots(1-aq^{n-1}).
  \label{defqPoch}
\end{equation}
$\circ$ hypergeometric function ${}_rF_s$ :
\begin{equation}
  {}_rF_s\Bigl(\genfrac{}{}{0pt}{}{a_1,\,\cdots,a_r}{b_1,\,\cdots,b_s}
  \Bigm|z\Bigr)
  \eqdef\sum_{n=0}^{\infty}
  \frac{(a_1,\,\cdots,a_r)_n}{(b_1,\,\cdots,b_s)_n}\frac{z^n}{n!}\,,
  \label{defhypergeom}
\end{equation}
where $(a_1,\,\cdots,a_r)_n\eqdef\prod_{j=1}^r(a_j)_n
=(a_1)_n\cdots(a_r)_n$.\\
$\circ$ $q$-hypergeometric function (the basic hypergeometric function)
${}_r\phi_s$ :
\begin{equation}
  {}_r\phi_s\Bigl(
  \genfrac{}{}{0pt}{}{a_1,\,\cdots,a_r}{b_1,\,\cdots,b_s}
  \Bigm|q\,;z\Bigr)
  \eqdef\sum_{n=0}^{\infty}
  \frac{(a_1,\,\cdots,a_r\,;q)_n}{(b_1,\,\cdots,b_s\,;q)_n}
  (-1)^{(1+s-r)n}q^{(1+s-r)n(n-1)/2}\frac{z^n}{(q\,;q)_n}\,,
  \label{defqhypergeom}
\end{equation}
where $(a_1,\,\cdots,a_r\,;q)_n\eqdef\prod_{j=1}^r(a_j\,;q)_n
=(a_1\,;q)_n\cdots(a_r\,;q)_n$.

\bigskip
\noindent{\bf AUTHOR DECLARATIONS}

\noindent{\bf  Funding}

No funds, grants or other support was received.

\noindent{\bf Conflict of Interest}

 The author has no conflicts to disclose.

\noindent{\bf DATA AVAILABILITY}

 Data sharing is not applicable to this article as no new data were created or analyzed in this study.


\end{document}